\documentclass[11pt]{article}
\pdfoutput=1
\usepackage{jcappub}

\usepackage{graphicx, color}

\usepackage{bm}
\usepackage{aas_macros}

\newcommand{\half}{\frac{1}{2}}

\newcommand{\cholD}{\mathsf{D}}
\newcommand{\cholT}{\mathsf{T}}
\newcommand{\matC}{\mathsf{C}}

\newcommand{\cholL}{\mathsf{L}}
\newcommand{\matPhi}{\mathsf{\Phi}}

\newcommand{\thetab}{\boldsymbol\theta}
\newcommand{\phib}{\boldsymbol\phi}
\newcommand{\mub}{\boldsymbol\mu}
\newcommand{\gammab}{\boldsymbol\gamma}
\newcommand{\deltab}{\boldsymbol\delta}

\newcommand{\ident}{\ensuremath{\mathbb{I}}}

\title{On estimating cosmology-dependent
covariance matrices}

\author[a]{Christopher B. Morrison}
\author[b,a]{and Michael D. Schneider}

\affiliation[a]{University of California, Davis, 
One Shields Ave., Davis, CA, 95616, USA.}
\affiliation[b]{Lawrence Livermore National Laboratory, \\P.O. Box 808 L-210, Livermore, CA 94551-0808, USA.}

\emailAdd{cbmorrison@ucdavis.edu}
\emailAdd{schneider42@llnl.gov}

\subheader{LLNL-JRNL-635596}

\abstract{
We describe a statistical model to estimate the 
covariance matrix of matter tracer two-point 
correlation functions with cosmological
simulations. Assuming a fixed number of cosmological 
simulation runs, we describe how to build a `statistical emulator' 
of the two-point function covariance 
over a specified range of input cosmological parameters. 
Because the simulation runs with different cosmological 
models help to constrain the form of the 
covariance, we predict that the cosmology-dependent 
covariance may be estimated with a comparable number of 
simulations as would be needed to estimate the covariance 
for fixed cosmology. 
Our framework is a necessary first step 
in planning a simulations campaign for analyzing the 
next generation of cosmological surveys.
}


\begin{document}
\bibliographystyle{JHEP}

\maketitle

\label{firstpage}

\section{Introduction}
\label{sec:introduction}

Cosmological large-scale structure statistics contain valuable 
information about many cosmological parameters. 
Accurate parameter inference from large-scale structure observations 
requires a model for the sample variance distribution of the 
observed statistics, for example the uncertainties and correlations 
between angular scales in the galaxy two-point correlation function.
Often, the sample variance distribution is assumed to be 
multivariate Gaussian, requiring specification of only a covariance
matrix of the cosmological two-point functions.
While the covariance of the two-point function is known
for a Gaussian field, effects of survey masks, galaxy clustering bias, 
and nonlinear gravitational evolution require many simulated 
realizations of the survey via $N$-body codes to accurately predict the covariance
structure. 
Ref.~\cite{dodelson13} recently showed that errors in covariance 
estimates propagate into increased cosmological parameter errors, which 
have the same effect as a reduction in the survey area. Near-term 
surveys will require at least $10^{4}$ simulation realizations 
to prevent effective survey area losses larger than 10\%.
But, the computational requirements for covariance estimation
will be even more challenging than forecasted in Ref.~\cite{dodelson13}
because the cosmology dependence of the covariance is 
likely to become important
as cosmological parameter constraints shrink in future surveys~\cite{eifler09, jee13}.

In some analyses, the covariance matrix was estimated from the 
data using resampling methods such as the Jackknife or bootstrap.
This has two large disadvantages. First,~\cite{norberg08} 
showed that a variety of resampling methods underestimate both 
the variances and correlations compared to those derived from 
many simulation realizations. Second, any data-derived 
covariance estimator necessarily ignores the cosmology-dependence
in the covariance which can significantly bias results~\cite{eifler09}.

In this paper we aim to determine how many cosmological simulations 
are required to achieve a target uncertainty in the covariance 
of the two-point correlations of mass density tracers.
We also show how the number of simulation runs can be reduced over the brute 
force approach by  
exploiting the smooth variation of the covariance components as a 
function of standard cosmological parameters in a simulation emulator.

We briefly argue for the need to model the cosmology dependence of the 
covariance in Section~\ref{sec:covariance_impact}, although 
this has also been established in the literature~\citep{eifler09, jee13}. 
In Section~\ref{sec:covariance_emulator} we specify the statistical 
model for the cosmology-dependent covariances that allows estimation
of covariances with a minimal number of simulations.
In Section~\ref{sec:simulation_design_study} we apply 
the Fisher matrix to forecast the uncertainties in covariance matrix 
elements as functions of the number of cosmological simulations run.
We summarize our main conclusion in Section~\ref{sec:conclusions}
that it is more efficient to simultaneously model the CDC 
than run many simulation realizations at several fixed
cosmological models. We also briefly describe the halo model used
in this analysis and assess its accuracy the Appendix.

\section{Why cosmology-dependent covariances?}
\label{sec:covariance_impact}

To motivate a two-point statistic covariance emulator, it is helpful to understand 
how cosmological parameter inference is affected by the 
cosmology dependence of the covariance.
First, there is an issue of data interpretation: When comparing
a model to data is it more helpful to ascribe uncertainties to the
data (i.e. from observational limitations) or the model (i.e. 
from the statistical formulation of the statistics that are 
observed, commonly referred to as `sample variance')?
Second, we should determine how biased our parameter constraints
could be if we erroneously ignore the cosmology dependence of the 
covariance. This latter issue was already addressed in~\cite{eifler09} who showed that the inferred parameter constraints can change by many standard deviations in some cosmic shear scenarios and can also be biased. We leave similar predictions using our covariance emulator to future work.

As stated previously, estimating the covariance model from the data tends to 
underestimate the uncertainties~\cite{norberg08} when compared with ensembles of simulated observations.
Simulating the error on the model is then the preferred method, however, 
the model error is itself often model dependent. 

For example, consider $N$ observed samples $x_i$ from a one-dimensional 
zero-mean Gaussian distribution where we aim to estimate the 
standard deviation of the distribution. The maximum likelihood 
estimator for the standard deviation is,
\begin{equation}
    \hat{\sigma}^2 = \frac{1}{N}\sum_{i} x_{i}^{2}.
\end{equation}
This estimator has an error,
\begin{equation}
    {\rm var}\left(\hat{\sigma}^2\right) =
    \frac{(2N-1)\sigma^2}{N^2},
\end{equation}
so the error on the estimator is also dependent on the value 
of the model parameter $\sigma^2$.
This is a simple example, but the same principle holds for 
many cosmological estimators.

\begin{figure}
    \centerline{
        \includegraphics[width=0.9\textwidth]{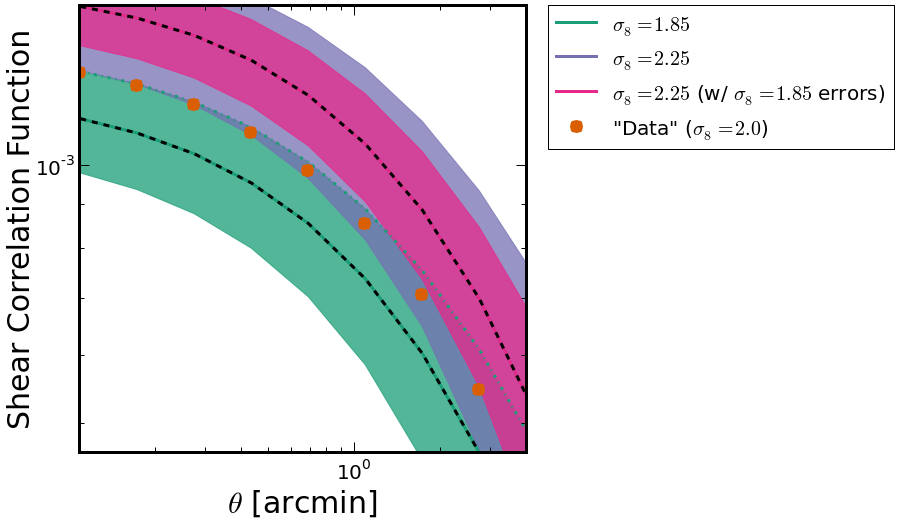}
    }
    \caption{A cartoon example illustrates how parameter-dependent 
    model errors can be important. A given set of observations,
    orange points, may be well fit by a model with a small amplitude 
    and smaller model errors (green shaded region) or by a larger
    amplitude model that also has larger model errors (purple
    shaded region). If the errors from the model with the smaller 
    amplitude are erroneously assumed to hold for all models we get 
    the magenta band around the higher amplitude model; leading 
    to a worse model fit to the orange data points.}
    \label{fg:sigma8_example}
\end{figure}

We show an example of parameter-dependent errors on the 
angular correlation function in Figure~\ref{fg:sigma8_example} when 
the parameter is the square-root of the 
amplitude of the two-point function, $\sigma_8$.
In this example, the data points 
are generated with $\sigma_{8}=2$ (for illustration only) and are consistent 
with models that have 
$\sigma_8=1.85$ or $2.25$
when the proper error is used. 
However, if one were to erroneously assign the errors for $\sigma_8=1.85$ to models 
with varying $\sigma_8$ in the mean correlation function (i.e. a model-independent error), 
the model with the
larger true variance would be incorrectly excluded by the data.
This situation always arises when assuming
fixed errors for cosmological models with different $\sigma_8$ (although 
at a less obvious level for observationally consistent values of $\sigma_8<1$).
The same principle applies to inference of other cosmological parameters 
as well.

\section{Covariance matrix emulator and simulation design} 
\label{sec:covariance_emulator}

The covariance matrices of large-scale 
structure probes, $\matC$ 
are typically estimated by running many $N$-body simulations with
different pseudo-random number seeds in the initial conditions and 
constructing a sample covariance estimator from the outputs. This is 
a computationally intensive task, requiring~$10^4$~$N$-body 
simulations to reduce the errors in the covariance below other systematic uncertainties for current surveys and several fold for surveys of the 
future. These simulations must also be run for different cosmologies 
in order to properly give unbiased constraints. We address both of 
these issues with a statistical emulator described in this section that allows for 
a reduction in the number of simulations required as well as linking 
together the simulations run at different cosmologies.

We consider $N_r\equiv \sum_{i=1}^{n_d} n_{r,i}$ simulations run at 
$n_d$ points in parameter space with $n_{r,i}$ independent simulation realizations 
at each point. For each simulation a summary statistic $\mathbf{y}^{i}_{k}$
(such as the power spectrum or correlation function)  is computed 
(where $k=1,\dots,n_{r,i}$). Given the set of $\mathbf{y}^{i}_{k}$,
statistics we now describe how to estimate a model for the 
cosmology-dependent covariance (CDC) by means of a `simulation emulator'.

Simulation emulators have been successfully developed 
to model the mean matter power spectrum over a six-dimensional
cosmological parameter space at high precision~\citep{lawrence10} 
and have been applied successfully to problems where the input parameters are less well-constrained~\cite{bower10, lu11, lu12, gomez12}. 

An emulator is specified in 2 steps. 
The `simulation design' defines at which points in 
cosmological parameter space $N$-body simulations will be run to 
calibrate the emulator.
The `emulation' step consists of calibrating a statistical model to 
interpolate the outputs of the simulation design runs to new points 
in parameter space.

The first step in building a simulation design is choosing the points 
in parameter space where simulations will be run. The Orthogonal 
Array Latin Hypercube has proven to be a successful algorithm 
for choosing design points (but see also the improvements in 
\cite{schneider11a}). 

Given specified design points, we then have to decide how many 
simulation realizations to run at each design point. 
We could require many simulation realizations at each design point 
if we need to construct a converged sample covariance estimator. 
However, with a careful parameterization of the covariance, we 
can use the simulation runs at all design points simultaneously 
to jointly constrain the covariance model at each design point.

We achieve our goal in several steps with four features of the
covariance emulator that all contribute to improve the estimate of 
the CDC over a brute-force sample covariance estimator,
\begin{enumerate}
\item a careful {\it decomposition} of the two-point covariance matrix,
\item a specification of orthogonal {\it basis functions} to decorrelate the 
covariance's components,
\item {optimized \it Gaussian Process parameters} derived from the ensemble of simulation runs,
\item calibration of the emulator to constrain the decorrelated {\it mode amplitudes of the covariance matrix components}.
\end{enumerate}
Briefly, we first decompose our matrices using the Generalized Cholesky 
Decomposition to produce values that are easily emulated. Second, 
we create basis functions from these values using Principal Component
Analysis (PCA). Third we fit (via maximum-likelihood) 
Gaussian Process parameters to link together
the different cosmologies in the simulation design. Finally, we
constrain our PCA mode amplitudes using the Gaussian Process.

An important consideration that we neglect in this paper is whether 
a (potentially noisy) sample covariance estimator is first needed 
at each design point before the covariance emulator can be constructed. 
We avoid this issue by using an analytic model for the covariances, 
but Ref.~\cite{schneider08} showed that sample covariance estimates might not be
needed at any stage in the calculation. We will explore this further 
in a later publication.

\subsection{Covariance matrix decomposition}
\label{sub:simulation_design}

We use the Generalized Cholesky Decomposition (GCD) to decompose the 
covariance matrices~\cite[following][]{pourahmadi07, daniels09, pourahmadi11}, 
which is alternately written in 
the following forms,
\begin{align}
  \matC &= \cholL \cholD \cholL^{T} \\
  \matC^{-1} &=  \cholT^{T} \cholD^{-1} \cholT, \qquad
  \cholT \equiv \cholL^{-1},
\end{align}
where $\cholT$ is a lower-triangular matrix with ones 
on the diagonal and 
$\cholD$ is a diagonal semi-positive definite matrix. 
The primary utility of the GCD for our purposes is that a 
positive definite 
covariance matrix, $\matC$, is guaranteed as long as all diagonal entries 
of $\cholD$ are positive, for any real values $\phi$ in the lower triangular components 
of $\cholT$. We can therefore interpolate unconstrained values of $\phi$ and $\ln\cholD$ 
over the cosmological parameter design space.

We will index the GCD components at each simulation design 
point with $i=1,\dots, n_d$.
Assuming the two-point function has $n_b$ bins, we label the
non-trivial components of $\cholT$ as~\cite{pourahmadi07},
\begin{equation}
  \cholT^{i}_{j\ell} = -\phi_{jl}^{i}
  \quad {\rm for}\, 2\le j \le n_b,\, \ell=1,\dots,j-1.
\end{equation}
Similarly, we label the diagonal components of $\cholD$ as,
\begin{equation}
  \cholD_{jj}^{i} = \exp(d^{i}_{j})
  \quad {\rm for}\, 1 \le j \le n_b. 
\end{equation}

\subsection{Basis functions}

The GCD provides a method to separate the $n_b(n_b+1)/2$ unique components of the 
covariance matrix into a computationally convenient form. 
We can further improve the covariance estimation by reducing the number of 
components that must be modeled as functions of cosmology. 
At the same time, we can use all the simulation runs, with all cosmological inputs, 
to determine common structures in the covariance matrices.

Following the emulator construction in~\cite{schneider08, lawrence10}, 
we proceed by extracting the components of $\ln\cholD$ and $\cholT$ into separate vectors for 
each design point, stack these vectors for all design points into an $n_y\times n_d$ 
array (where $n_y$ is either equal to $n_b$ or $n_b(n_b-1)/2$), and compute Principal Components (PC) to identify a subset of components that 
are most strongly varying over the design space.
We choose to compute principle components of $\ln\cholD$ so that 
we obtain values supported on the entire real line.

Following~\cite{schneider11a} we first subtract the
mean of the design run components $d$ and $\phi$  
and then scale the result by a single number so that the
combined entries of our $n_y \times n_d$ matrix of design run components have
variance one.  
\begin{align}
    \mathbf{d}^{i} &= \sigma_{d}\tilde{\mathbf{d}}^{i} + \mathbf{d}_{\rm C}
    \\
    \phib^{i} &= \sigma_{\phi}\tilde{\phib}^{i} + \phib_{\rm C}.
\end{align}
We collectively label the centered and scaled matrices by $\mub \equiv \tilde{\mathbf{d}}$ or 
$\tilde{\phib}$, and perform a singular value decomposition: $\mub=\mathsf{UBV}^T$
where $\mathsf{U}$  
has dimension $n_{y}\times p$ ($p \equiv {\rm min}(n_{y},n_{d})$) with 
$\mathsf{U}^{T}\mathsf{U}=\ident_{p}$, $\mathsf{V}$ 
has dimension $n_{d}\times p$ with $\mathsf{V}^{T}\mathsf{V}=\ident_{p}$,
$\mathsf{V}\mathsf{V}^{T}=\ident_{n_{d}}$, and $\mathsf{B}$ ($p\times
p$) is a diagonal matrix of singular values.  The
matrix of basis vectors, $\mathsf{\Phi}\equiv\frac{1}{\sqrt{n_{d}}}\mathsf{U}\mathsf{B}$
with weights $\boldmath{w}\equiv\sqrt{n_{d}}\mathsf{V}^{T}$ normalized so that
$\frac{1}{n_{d}}\boldmath{w}^{T}\boldmath{w}=\ident_{n_{d}}$ (this latter choice 
makes it simple to specify priors on the Gaussian Process variance parameters). 

We keep only the first $p_{D}, p_{\phi} \le p$ columns of $\mathsf{\Phi}$.
Then we redefine the scaled design covariance matrix components as a sum 
over $p_{D}$ or $p_{\phi}$ modes,
\begin{align}\label{eq:modedecomp}
  \tilde{d}_{t}(\thetab_i)  &= 
  \sum_{j=1}^{p_{D}} \gamma_{j}^i \Phi_{D, t j},
  \qquad t = 1,\dots,n_b
  \notag\\
  \tilde{\phi}_{t}(\thetab_i)  &= 
  \sum_{j=1}^{p_{\phi}} \delta_{j}^i \Phi_{\phi, t j},
  \qquad t = 1,\dots,\half n_b(n_b-1)
\end{align}
where $\Phi_{X, t j}$ is the entry of the matrix $\mathsf{\Phi}_{X}$ in row $t$
($t=1,\dots,n_y$) and column $j$ ($j=1,\dots,p_{D}$ or $p_{\phi}$), 
$\gamma_{j}(\thetab)$ and $\delta_{j}(\thetab)$ 
are the $j$th
(parameter-dependent) mode amplitudes ($\sqrt{n_d}\mathsf{V}^{T}$ above),
and $\thetab$ are the model (i.e. cosmological) parameters.
Ref.~\cite{schneider08} included i.i.d. Normal errors in the truncated mode 
decompositions to account for residuals when $p_{D,\phi} < p$ (which will always 
be assumed so that the PCA achieves some reduction in parameters). 
For the purposes of forecasting we ignore the error in the truncation of the 
PCA expansion, but this error should be propagated when constructing 
a full emulator of the covariance (see e.g. Eqs. 7-11 of \cite{schneider08}).

Note that~\cite{pourahmadi07} suggest a similar decomposition for the 
simultaneous modeling of several covariance matrices. Our 
Eq.~\ref{eq:modedecomp} differs in using basis vectors $\mathsf{\Phi}$ 
that are independent of the model parameters $\thetab$ while imparting 
all the model dependence to the mode amplitudes in the truncated basis.

Joint estimation of the covariance matrices is now reduced to estimating the 
mode amplitudes $\gamma_{j}^{i}$ and $\delta_{j}^{i}$ from the 
simulations run at each design point, given the 
basis vectors $\mathsf{\Phi}_{D}$ and $\mathsf{\Phi}_{\phi}$, which 
are estimated from the combination of all simulation runs.

We define estimators for the mode amplitudes from the likelihood for the 
simulation design runs given the model in Eq.~\ref{eq:modedecomp}.
Ref.~\cite{pourahmadi07} show that with the GCD, the log-likelihood for 
the set of covariance matrices at the simulation design points can 
be written,
\begin{equation}\label{eq:likelihood}
  \ell \left(\mathbf{y}^{i}_{k}\right | \gamma^{i}_{j}, \delta^{i}_{j}) =
  \sum_{i=1}^{n_d} \sum_{j=1}^{n_b}
  \left(
  -\frac{n_{r,i}}{2} d_{j}^{i} - 
  \frac{n_{r,i}}{2} \hat{T}_{i,j}^{T} S_{i} \hat{T}_{i,j} \exp(-d_{j}^{i})
  \right),
\end{equation}
where $\hat{T}_{i,j}$ is the $j$th column of $\cholT_{i}$ and 
\begin{equation}
    S_{i} \equiv \frac{1}{n_{r,i}} \sum_{k=1}^{n_{r,i}}
    \left(\mathbf{y}^{i}_{k} - \bar{\mathbf{y}}^{i}\right)
    \left(\mathbf{y}^{i}_{k} - \bar{\mathbf{y}}^{i}\right)^{T}
\end{equation}
is the sample covariance matrix estimate at design point $i$. 
The mean $\bar{\mathbf{y}}^{i}$ could either be specified 
by a theoretical model (e.g. {\tt halofit}) or could be the sample mean 
from the simulation realizations at each design point. 
Ref.~\cite{schneider08} consider how the sample mean can 
be jointly estimated with the sample covariance, but here 
we assume the mean has zero uncertainty.

We can then estimate the mode amplitudes conditioned on the 
simulation design runs either with maximum-likelihood estimators 
or posterior samples.
In order to use simulation runs with different input cosmologies 
to jointly constrain a covariance model, we now must 
specify how the mode amplitudes $\gamma_{j}(\thetab)$ and 
$\delta_{j}(\thetab)$ can vary with $\thetab$.

\subsection{Parameters for the Gaussian Process}
\label{sub:gaussian_process_priors}

\newcommand{\yb}{\mathbf{y}}
\newcommand{\ab}{\mathbf{a}}
\newcommand{\alphab}{\boldsymbol\alpha}
\newcommand{\rhob}{\boldsymbol\rho}

We impose Gaussian Process (GP) priors on the mode amplitudes 
$\gamma^{i}_{j}$ and $\delta^{i}_{j}$ as a means of linking the covariance
models with different input cosmologies (in addition to the 
common structure imposed by the basis vectors 
$\mathsf{\Phi}_{D}$ and $\mathsf{\Phi}_{\phi}$)
as well as a 
means of interpolating the covariance predictions to regions 
of parameter space where no simulations 
have been run~\citep{habib07, schneider08, lawrence10, schneider11a}.
The GP priors impose requirements on the smoothness of the 
mode amplitude surfaces over the cosmological parameter space.

We model each mode amplitude as independent GPs with zero mean (because we 
already subtracted the mean values of the $d$ and $\phi$ over the simulation 
design) and an exponential covariance model described by a single precision 
parameter for each mode $\lambda_{X,j}$ and correlation parameters for 
each direction in the $p_{\theta}$-dimensional cosmological parameter
space $\rho_{X,ij} \in [0,1]$~\cite{schneider08},
\begin{equation}\label{eq:gpcov}
    \Sigma(\thetab, \thetab'; \rhob_{X,j}, \lambda_{X,j}) = 
    \lambda_{X,j}^{-1}\prod_{i=1}^{p_{\theta}} 
    \rho_{X,ij}^{4(\theta_{i}-\theta_{i}')^2},
\end{equation}
where $X=D,\phi$ and $j=1,\dots,p_{D,\phi}$.
The correlation parameters control the smoothness of the mode amplitudes 
along each parameter direction. If all $\rho_{X,ij}\approx1$ then the 
modes of the covariance are highly correlated over the design parameter 
space, allowing all simulations to aid in the estimation of the joint 
covariance model. Conversely, when many of the correlation parameters 
are near zero, we expect the covariance emulator to give little 
improvement beyond the standard sample covariance estimators at 
each design point. 

Our argument for using the covariance model in Eq.~\ref{eq:gpcov} rather 
than some other model is simply because it works, as we show in
Section~\ref{sec:simulation_design_study}.  
The Mat\'{e}rn covariance is a more flexible covariance model that is 
often used for GPs. Our choice of the covariance model in Eq.~\ref{eq:gpcov} is 
partly informed by our expectation that the cosmological covariances 
we are interested in will be smoothly varying over the parameter spaces that 
are already tightly constrained by existing CMB and large-scale structure observations.
It is possible that parameters in alternative cosmological models (e.g. modified gravity
or dynamical dark energy) will be less constrained and a more flexible 
GP covariance model could be appropriate. 

Restricted to the design points $\thetab^{*}$, the priors on the design mode amplitudes are 
then multivariate Normal,
\begin{align}\label{eq:mode_amp_priors}
    \pi(\gammab_{j} | \lambda_{d,j}, \rhob_{d,j}) &= 
    N(0, \Sigma(\thetab^{*}; \rhob_{d,j}, \lambda_{d,j}))
    \\
    \pi(\deltab_{j} | \lambda_{\phi,j}, \rhob_{\phi,j}) &= 
    N(0, \Sigma(\thetab^{*}; \rhob_{\phi,j}, \lambda_{\phi,j})),
\end{align}
and the posterior for forecasting constraints on the mode amplitudes given the design 
runs is obtained by combining Eqs.~\ref{eq:likelihood} and \ref{eq:mode_amp_priors},
\begin{equation}\label{eq:posterior}
    p(\gammab_{j}, \deltab_{j} | \mathbf{y}^{i}, \lambda_{d,j}, \rhob_{d,j},
    \lambda_{\phi,j}, \rhob_{\phi,j}) =
    L\left(\mathbf{y}^{i}_{k}\right | \gamma^{i}_{j}, \delta^{i}_{j}) 
    \pi(\gammab_{j} | \lambda_{d,j}, \rhob_{d,j}) 
    \pi(\deltab_{j} | \lambda_{\phi,j}, \rhob_{\phi,j}).
\end{equation}
We later compute the Fisher matrix for the mode amplitudes 
by taking derivatives of Eq.~\ref{eq:posterior}.

\newcommand{\lambdab}{\boldsymbol\lambda}

The statistical model for the CDC is completed by calibrating 
the parameters of the GP models for each mode $\gammab_{j}$ and 
$\deltab_{j}$. So, the simulation design runs are used to both 
infer the structure of the covariance parameterization in the 
form of the basis vectors $\mathsf{\Phi}_D$ and $\mathsf{\Phi}_{\phi}$ 
and to infer the cosmology dependence in the form of the 
GP parameters for the mode amplitudes. 

Previous emulators in the cosmology literature built 
hierarchical Bayesian models to marginalize over the GP parameters
for each mode amplitude and thereby propagate all interpolation 
and parameterization uncertainties. We have built a similar 
framework here (in Eq.~\ref{eq:posterior}), 
but for the purposes of forecasting, 
will now use maximum-likelihood estimators for the GP parameters.

By definition, the likelihood for the Gaussian process model 
for a mode amplitude evaluated at the design points is 
multivariate Gaussian~\cite[eq. (5.8) of ][]{rasmussen06},
\begin{equation}
    \ln(p(\yb|\ab)) = 
    -\half \yb^{T} \Sigma_{y}^{-1} \yb 
    -\half \ln \left|\Sigma_{y}\right| 
    -\frac{n_d}{2}\ln 2\pi,
\end{equation}
where $\yb\equiv \left\{\delta_{j}^{i}; i=1,\dots,n_d\right\}$ or 
$\left\{\gamma^{i}_{j}; i=1,\dots,n_d\right\}$ as determined by our fiducial model,
and we have jointly labeled the GP model parameters as 
$\ab\equiv \left\{ \rhob, \lambda \right\}$.
Taking derivatives with respect to the GP parameters, 
\cite{rasmussen06} find (their eq. 5.9),
\begin{equation}
    \frac{\partial}{\partial a_i} \ln(p(\yb|\ab)) = 
    \half {\rm Tr}\left(
    \left(
    \alphab \alphab^T - \Sigma^{-1}
    \right)
    \frac{\partial \Sigma}{\partial a_{i}}
    \right)
    \qquad
    {\rm with}\,
    \alphab \equiv \Sigma^{-1}\yb.
\end{equation}

We also include hyperpriors on the parameters of the GP models,
\begin{align}
    \pi(\lambdab_{X}) &= \prod_{j=1}^{p_{X}} \lambda_{X,j}^{a_X-1}
    e^{-b_X \lambda_{X,j}}
    \\
    \pi(\rhob_{X}) &= \prod_{j=1}^{p_{X}} \prod_{i=1}^{p_{\theta}}
    \rho_{X,ij}^{a_{\rho, X}-1}
    \left(1 - \rho_{X,ij}\right)^{b_{\rho,X}-1}
\end{align}
So we use the same hyperprior parameters for all modes of $d$ and 
all modes of $\phi$.

\subsection{Mode amplitude constraints} 
\label{sub:emulation}

Next we use the Fisher matrix to forecast the errors on the 
components of the covariance matrix estimates at every 
point in our simulation design given the set of simulation 
design runs. We aim to determine how many cosmological simulations
are needed to achieve a given precision in the elements of the 
covariance matrix estimates given:
\begin{itemize}
  \item the number of simulation design points $n_d$,
  \item the number of simulation realizations 
  at each design point $n_{r,i}$,
\end{itemize}

A related question is whether it is better to estimate the 
high-precision 
covariances at a few fixed cosmological models, or to estimate 
the CDC jointly 
with a few realizations at many cosmological parameter values. 
The answer to this 
question will partly depend on how smoothly the chosen components 
of the covariance matrices vary over the cosmological parameter space. 
We will show that the GCD and PCA parameterization of the 
model covariances
described in section~\ref{sec:covariance_emulator} yield such smoothly 
varying parameters for a standard wCDM model.

The Fisher matrix is defined as the negative of the expectation of the curvature of the 
log-likelihood (or log-posterior as given in Eq.~\ref{eq:posterior}) about its peak,
\begin{equation}
  F_{ij} \equiv -\left< \frac{\partial^{2}\ell(\mathbf{y}, \thetab)}
    {\partial\theta_{i}\partial\theta_{j}}\right>_{\theta=\theta_0}.
\end{equation}
The terms contributing to the Fisher matrix for the CDC mode amplitudes are,
\begin{align}\label{eq:gammaFisherInfo}
  -\left<\frac{\partial^{2}\ell}{\partial\gamma^{j}_{k}\partial\gamma^{j}_{\ell}}\right>
  &= \half n_{r,j} \sigma_{d}^2 \Phi_{D,k}^{T} \Phi_{D,\ell}
  \notag\\
  &= \half n_{r,j} \frac{\sigma_{d}^2\, b^{2}_{j}}{n_d} \delta^{\rm D}_{k\ell}
\end{align}
where $\Phi_{k}$ is the $k$th row of the covariate matrix $\Phi$, $b_{j}$ is the 
$j$th singular value in the PC decomposition of the design runs, and we have 
used the definition of $\Phi$ as composed from the SVD of the design runs. So, the 
conditional Fisher information for the $\gamma$ mode amplitudes is the same 
for all design points, is independent for each mode amplitude, 
and only depends on the SVD of the 
design runs (in the form of the variance of each mode over the design). 
Eq.~\ref{eq:gammaFisherInfo} is analogous to the standard error on the variance.

For the $\delta$ modes however,
\begin{equation}\label{eq:deltaFisherInfo}
  -\left<
  \frac{\partial^{2}\ell}
  {\partial\delta^{i}_{\eta}\partial\delta^{i}_{\gamma}}
  \right>
  = n_{r,i} \sigma_{\phi}^2
  {\rm Tr}\left(
  {\rm tri}\left(\Phi_{\phi,\eta}\right)\, \matC^{i} \,
  {\rm tri}\left(\Phi_{\phi,\gamma}\right)^{T} \cholD^{-1}
  \right),
\end{equation}
where $\Phi_{\phi,\eta}$ is the column of $\matPhi_{\phi}$ indexed by $\eta$ 
and ${\rm tri}\left(\Phi_{\phi,\eta}\right)$ indicates we force 
$\Phi_{\phi,\eta}$ to fill the lower-triangular elements of an 
$n_b\times n_b$ matrix with all other entries zero.
So the Fisher information depends on the covariance 
$\matC^{i}$ at each design point $i$. And, 
the covariance between different $\delta$ modes 
is nonzero, which means that constraining some modes of 
$\phi$ (at a fixed design point) 
can help constrain the other modes as well. 

The joint Fisher matrix for $\gamma$ and $\delta$ has nonzero cross-terms,
\begin{equation}\label{eq:gammadeltaFisherInfo}
  -\left<
  \frac{\partial^{2}\ell}
  {\partial\delta^{i}_{\eta}\partial\gamma^{i}_{m}}
  \right>
  = n_{r,i} \frac{\sigma_{d}\sigma_{\phi}}{2}
  {\rm Tr}\left[
  {\rm tri}\left(\Phi_{\phi,\eta}\right)
  \left(\cholT^{i}\right)^{-1} 
  {\rm diag}\left(\Phi_{D, m}\right)
  +
  {\rm diag}\left(\Phi_{D, m}\right)
  \left(\cholT^{i}\right)^{-1} 
  {\rm tri}\left(\Phi_{\phi,\eta}\right)  
  \right]
\end{equation}

Notice that eqs.~\ref{eq:gammaFisherInfo}, 
\ref{eq:deltaFisherInfo}, and \ref{eq:gammadeltaFisherInfo}
scale linearly with $n_{r,i}$ as might be expected. 
From this we can immediately 
see that without a model for connecting the mode 
amplitudes at different design 
points, it is not possible to reduce the error in 
the covariance matrix 
elements faster than $\sqrt{n_{r}}$.

\section{Simulation design study: cosmic shear}
\label{sec:simulation_design_study}

\begin{figure}
    \centerline{
        \includegraphics[width=0.5\textwidth]{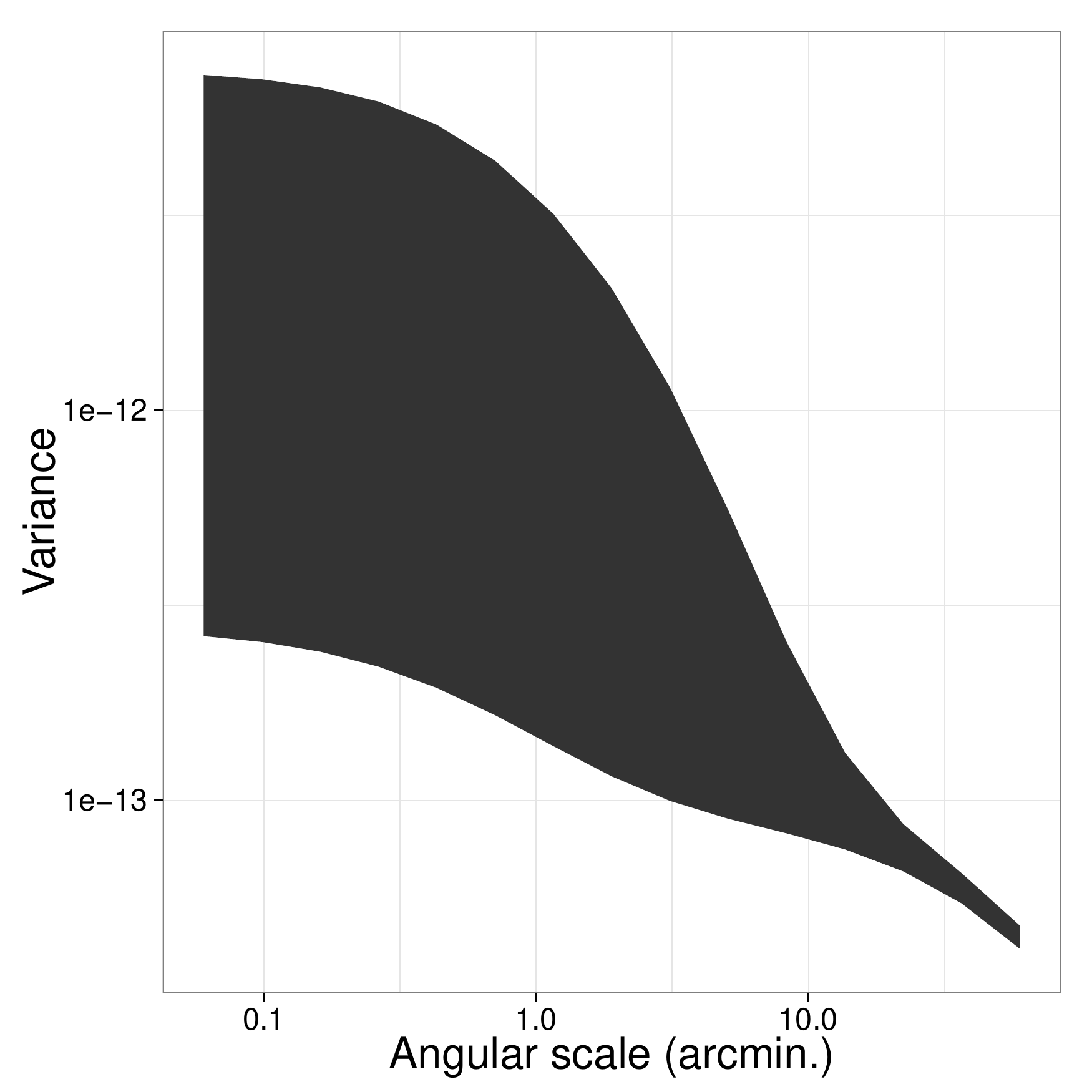}
        \includegraphics[width=0.5\textwidth]{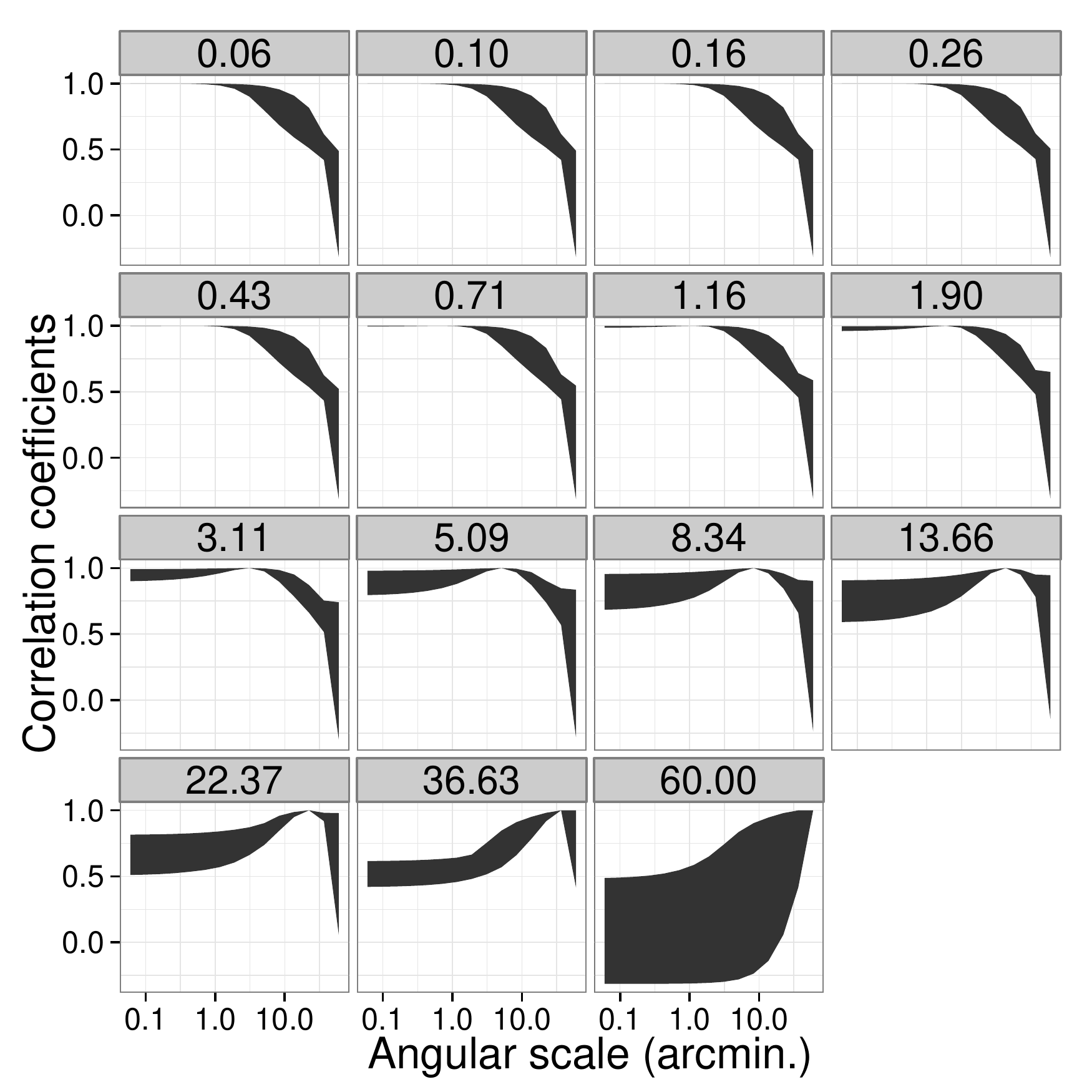}
    }
    \caption{The range of values of the covariance of the 
    shear correlation function over the simulation design space.
    Left: diagonal terms in the covariance. Right: rows of the 
    matrix of correlation coefficients. Each panel represents 
    a different bin in angular scale in the correlation function (in arcmin.). For
    the right pannel, we see the qualitative trends we expect, strong
    correlation on small scales and weaker at large scales. The bands with larger widths at
    arcminute scales is due in part to the change in angular 
    diameter distances with cosmology.}
    \label{fg:desvar}
\end{figure}

\begin{figure}
    \centerline{
        \includegraphics[width=0.5\textwidth]{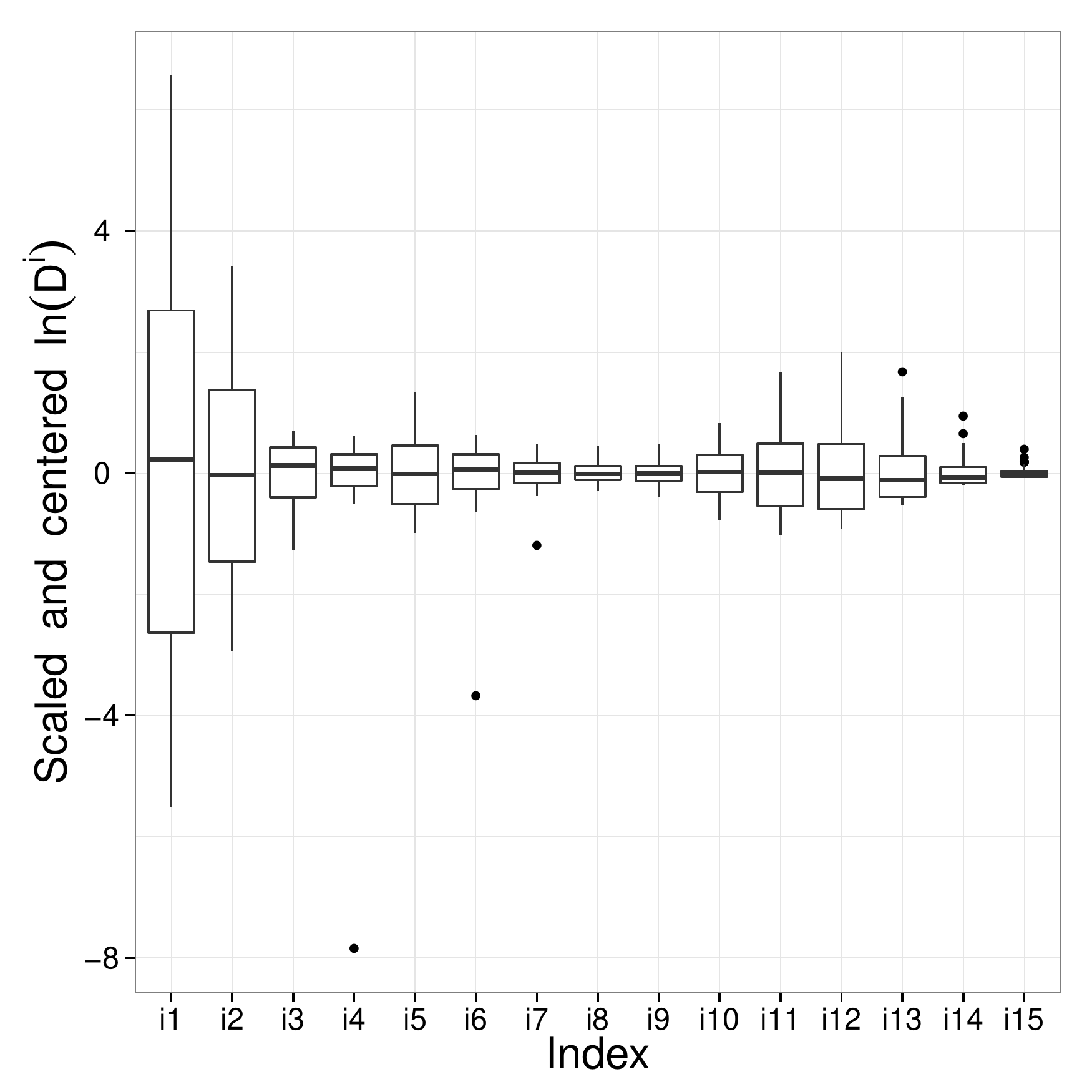}
        \includegraphics[width=0.5\textwidth]{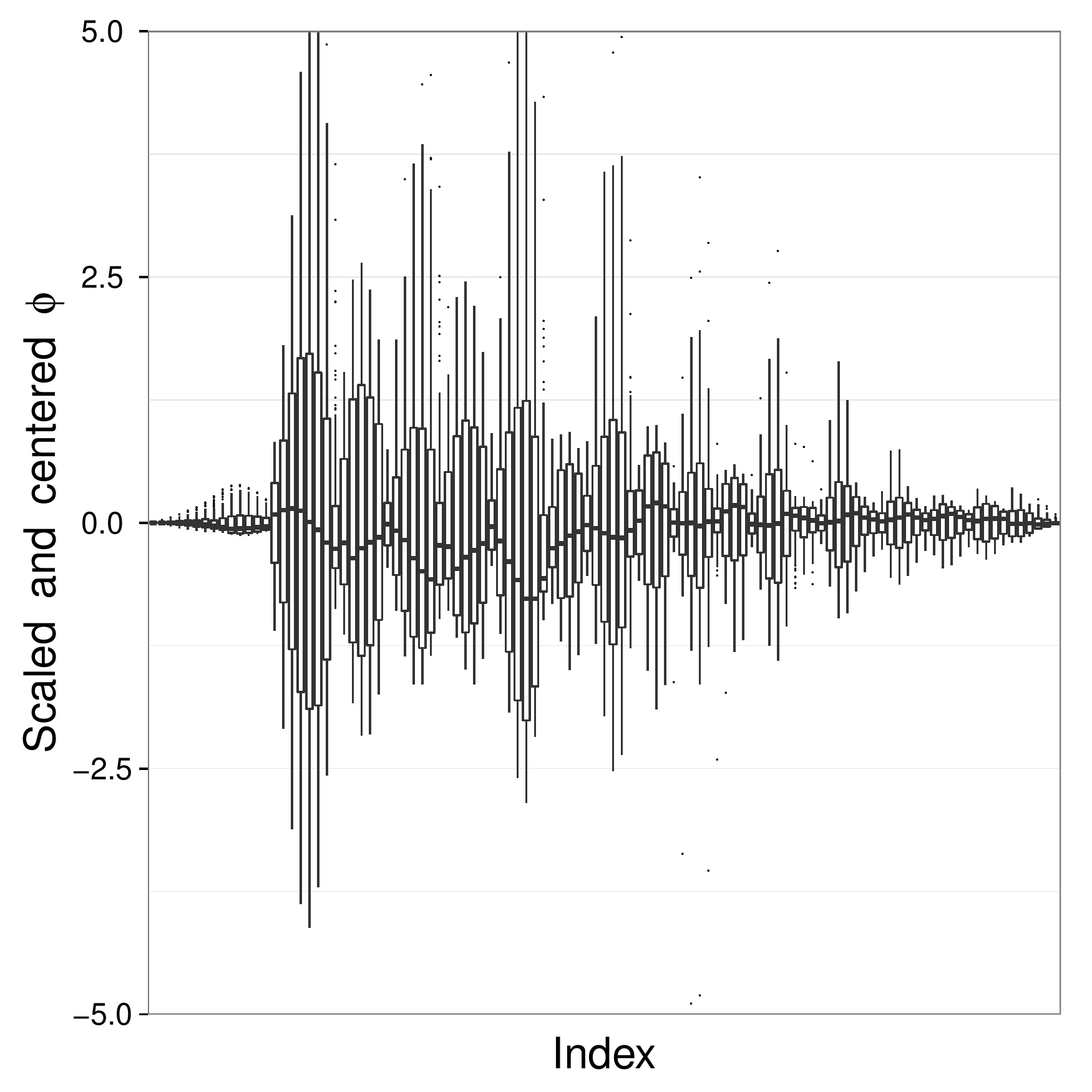}
    }
    \caption{Distribution of components of the Generalized Cholesky 
    Decomposition of the design covariance matrices for the different
    cosmology design points. In this decomposition, a majority of the
    variation in the matrix elements is confined to a few indices.
    }
    \label{fig:gchol_components}
\end{figure}

To asses the performance of the covariance matrix emulator, we use 
the analytical halo model~\citep{cooraysheth02} to predict 
covariance matrices of the shear correlation function. Previous 
studies have shown that the 
halo model captures the correct qualitative features of the 
nonlinear two-point function covariances. While our modeling 
is expected to lack precision relative to what would be 
estimated from $N$-body simulations, we believe the complexity 
of the model is sufficient to demonstrate the utility of our 
statistical framework. Also, any plans for running a large number of
simulations to estimate the CDC must initially rely on imperfect
models. For details on the model used see Appendix \ref{sec:halo_model}.

We assume the same cosmological parameters and similar ranges of variation as
in~Ref.~\cite{lawrence10}, shown in Table~\ref{tab:design_ranges}. 
All examples assume 32 design points (i.e. $n_d=32$) in an Orthogonal Array 
Latin Hypercube (OALH) spanning this 5-dimensional design space.
\begin{table}[htb]
\begin{center}
\begin{tabular}{ccc}
\hline
Parameter & Min. & Max. \\
\hline
$\sigma_8$ & 0.611 & 1.011 \\
$\Omega_m h^{2}$ & 0.119 & 0.31\\
$\Omega_b h^{2}$ & 0.0215 & 0.0235 \\
$n_s$ & 0.86 & 1.06 \\
$w$ & -1.3 & -0.7\\
\hline
\end{tabular}
\caption{\label{tab:design_ranges} Ranges of the cosmological parameters for our example OALH simulation design.}
\end{center}
\end{table}
At each of the 32 design points we compute a model for the 
nonlinear covariance of the shear correlation function assuming a $\delta$-function 
source distribution at $z=1$ and negligible shape noise. 

\begin{figure}
    \centerline{
        \includegraphics[width=0.9\textwidth]{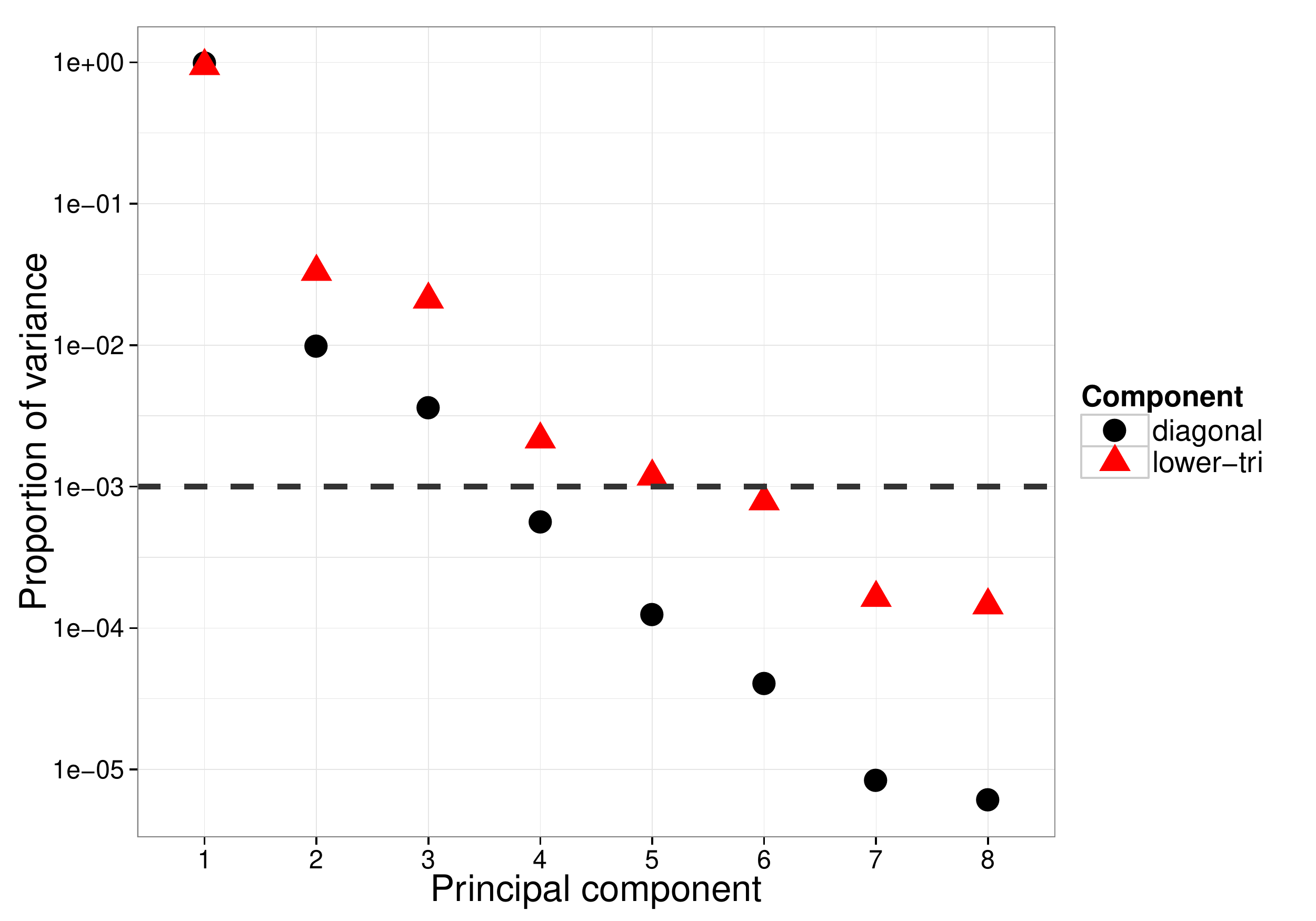}
    }
    \caption{The first 8 principal components of the simulation design 
    $\ln(D)$ (black circles) and $\phi$ (red triangles) 
    components of the Generalized Cholesky Decomposition of the shear correlation function 
    covariance matrix.
    We keep all PC modes that contribute a fraction of more than $10^{-3}$ to the 
    total variance (i.e. all modes above the horizontal dashed line).
}
    \label{fig:pca}
\end{figure}

\begin{figure*}
  \centerline{
    \includegraphics[width=0.5\textwidth]{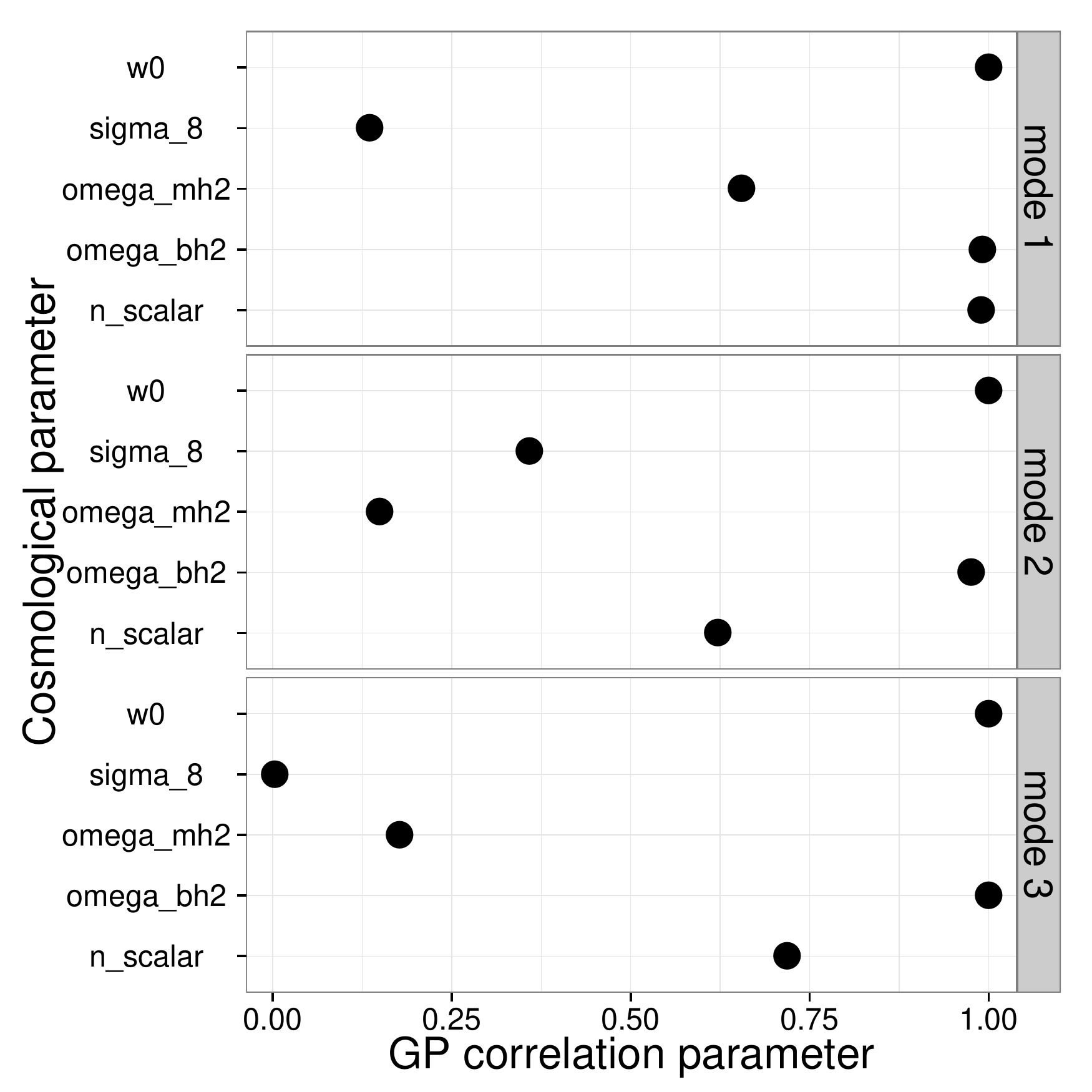}
    \includegraphics[width=0.5\textwidth]{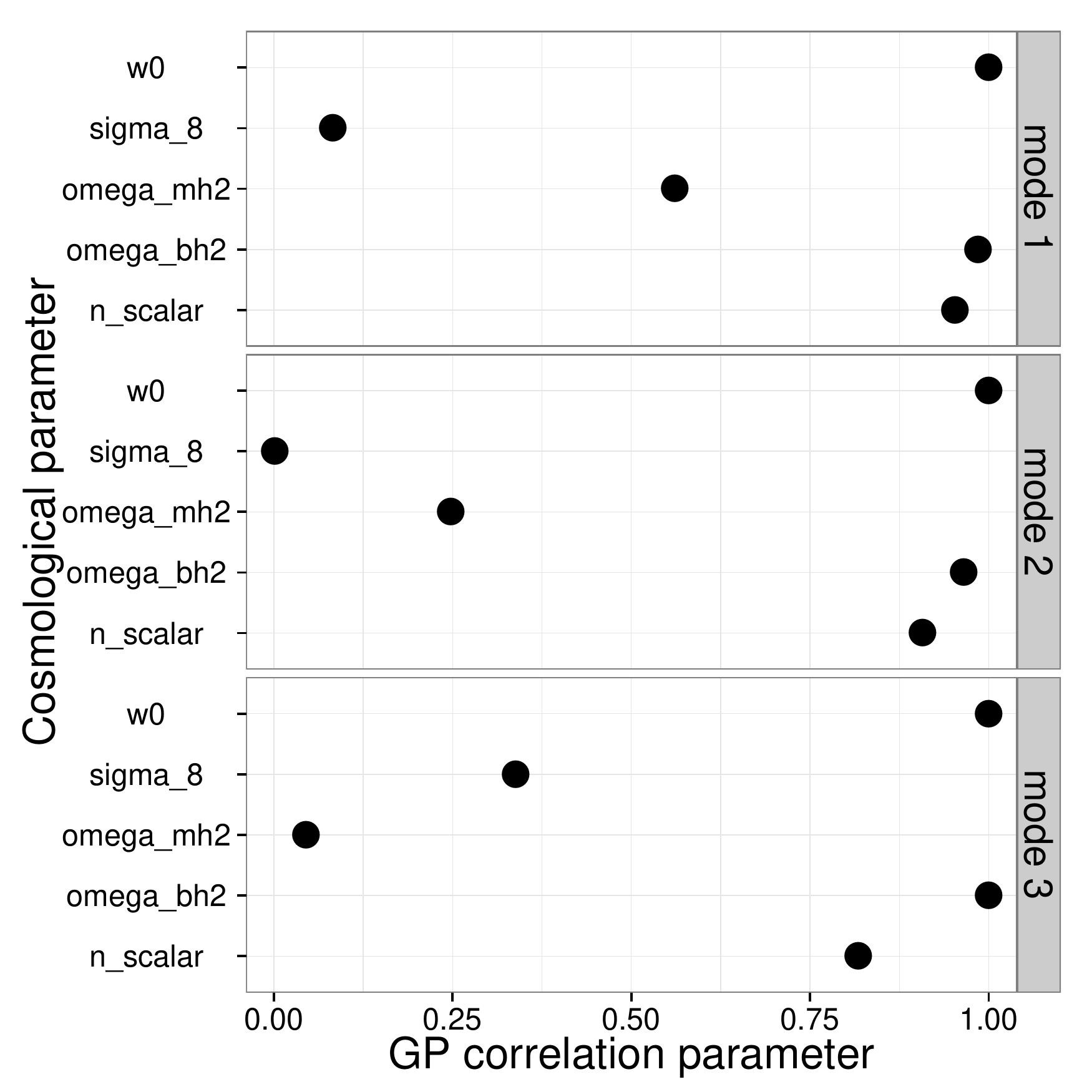}    
  }
  \caption{Maximum likelihood estimates for the GP correlation parameters 
  for the first 3 PC modes for the diagonal (left) and lower-triangular (right)
  components of the GCD of the covariance. 
  A correlation parameter close to one indicates 
  the mode amplitude is smoothly varying along a given parameter axis. Conversely,
  a correlation parameter near zero indicates large variations in the mode amplitude 
  along the given axis.}
  \label{fig:sensitivity}
\end{figure*}

\subsection{Covariance matrix decomposition}
In Figure \ref{fg:desvar} we show the full range of values in
covariance diagonal components (left) and correlation coefficients (right)
over the design space. The diagonal terms of the
covariance span well over an order of magnitude in value and differ
in shape as the angular projection of the one-halo term changes. The
amount of cross-correlation is mostly stable with cosmology however,
it does vary greatly at large scales. 
Figure \ref{fig:gchol_components} shows the amount of variation in
the GCD compoments, $\mathbf{d}$ and $\phib$ over the full simulated space.
In this decomposition we see much of the variation in the covariance
has been confined to fewer components.

\subsection{Basis functions}
In Figure \ref{fig:pca} we show the first 8 PCA amplitudes as a 
function of mode index. For both the diagonal elements $\mathbf{d}$ 
and lower triangular elements $\phib$ we see that most of the 
variation is  contained within the first few modes of the 
decomposition allowing for significant reduction in the 
dimensionality of the covariance matrices over the full simulation 
design. We retain those PC modes that contribute at least $10^{-3}$ 
to the fractional variance.

\subsection{Parameters of the Gaussian Process}
As stated previously in Section~\ref{sub:emulation} the amount of 
correlation in the GP modes informs us how strongly dependent the
information in the CDC is on different cosmological parameters.
Figure~\ref{fig:sensitivity} shows the mode amplitude
correlations (i.e. the $\rho_{X}$ parameters in Eq.~\ref{eq:gpcov}) 
with each cosmological parameter in a `sensitivity' analysis. 
A value of one means that the mode amplitudes are highly correlated 
along a given parameter axis, so the emulator is not sensitive to 
variations in this parameter. This also means sparser sampling of 
simulation design runs can be used.
The parameters $w_0$, 
$\Omega_{b}h^2$, and $n_s$ are often the most correlated, indicating they have 
little impact on the covariance. 
The mode amplitudes are consistently weakly correlated along the 
$\sigma_8$ and $\Omega_m h^2$ parameter axes, indicating these parameters 
largely determine the form of the covariance as might be expected because the 
amplitude of the shear correlation function depends on the product 
$\sim\sigma_{8}^2 \Omega_m^{0.6}$

\subsection{Mode amplitudes of the covariance matrices}

\begin{figure*}
  \centerline{
    \includegraphics[width=0.5\textwidth]{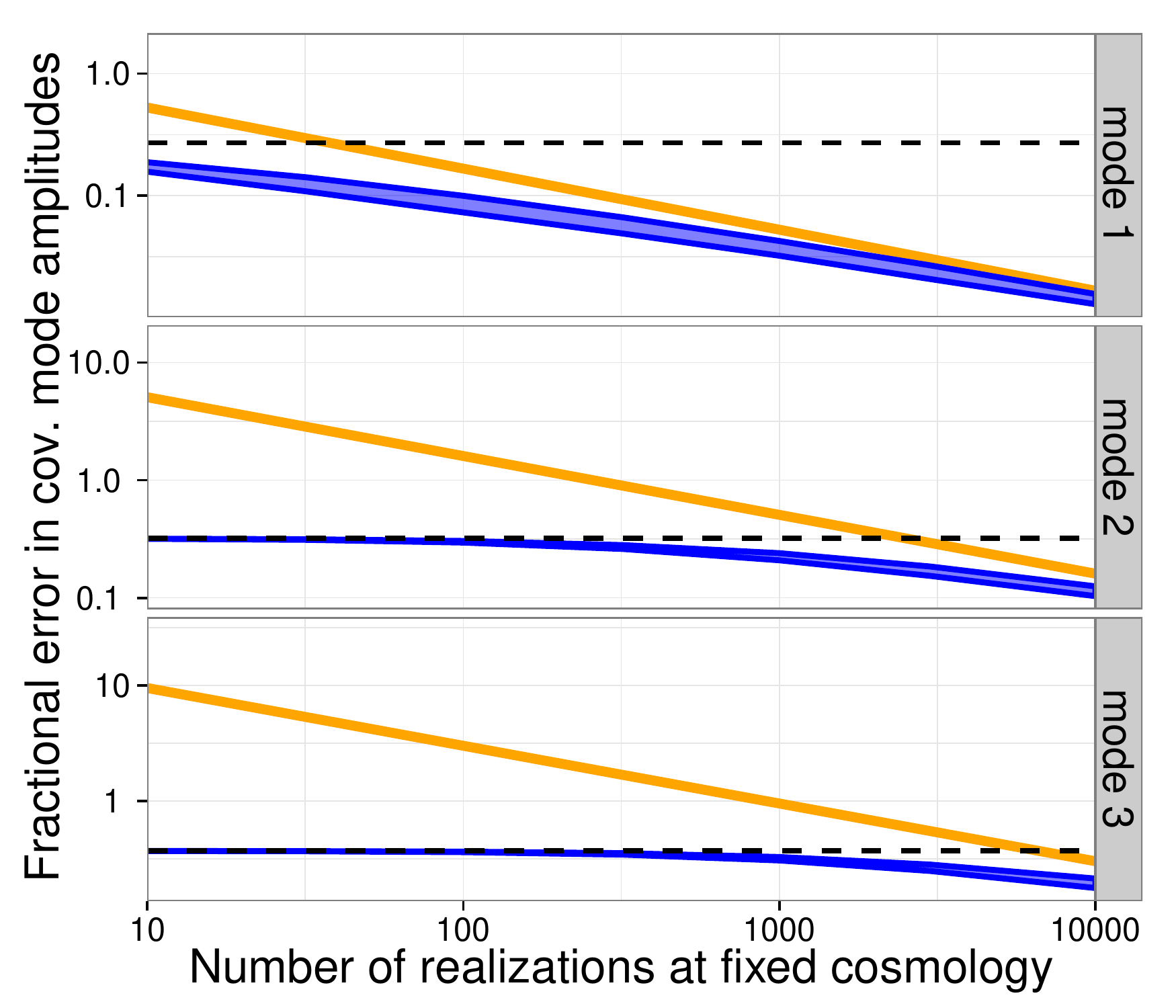}
    \includegraphics[width=0.5\textwidth]{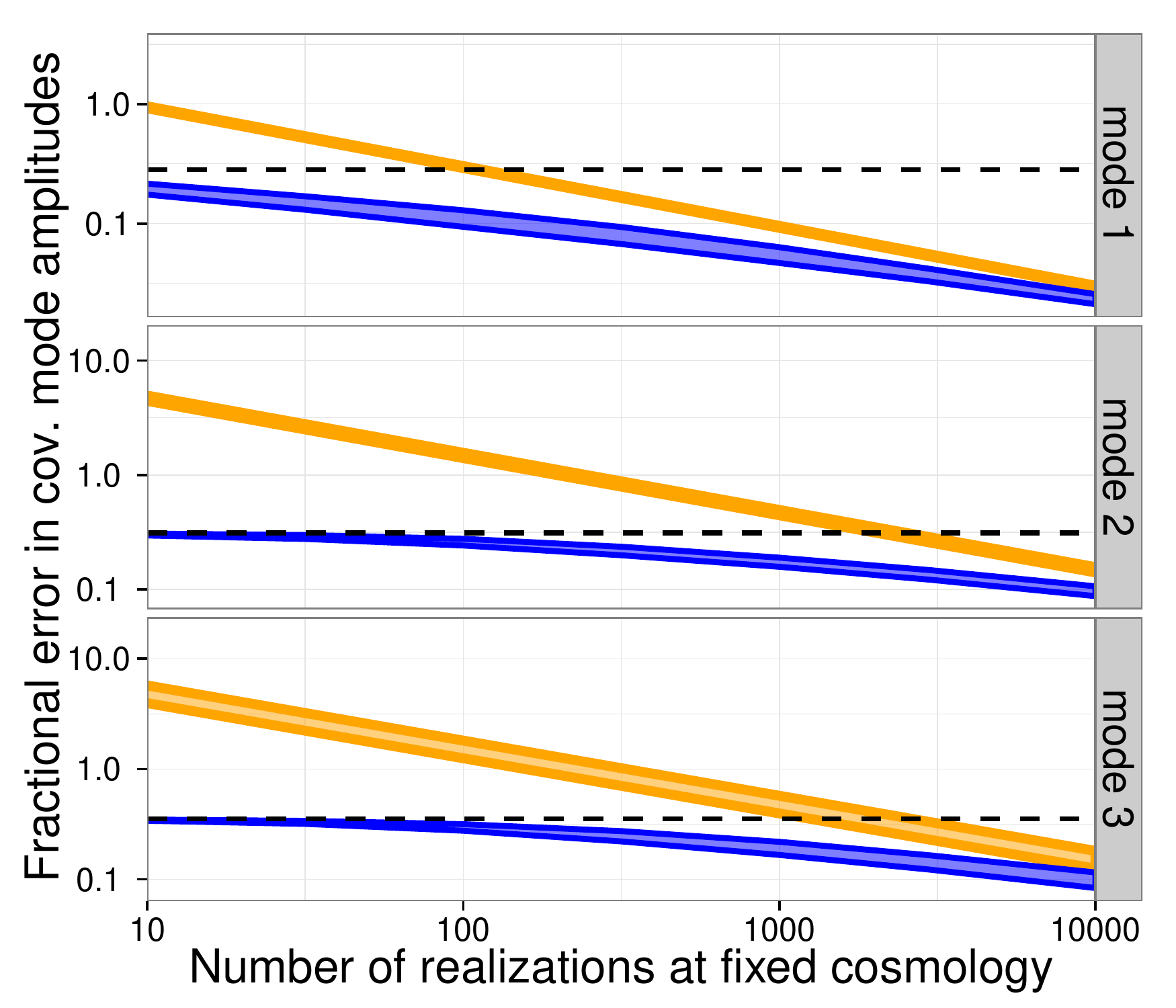}
  }
  \caption{Forecasted errors on the mode amplitudes of the 
  diagonal (left) and lower-triangular (right) components in the 
  GCD of the covariance matrices. The panels show the first 3 
  principle components in each case. The blue bands 
  show the range of marginal errors over 32 design points 
  when using all design 
  points simultaneously to constrain the mode amplitudes. 
  The orange bands show the range of marginal errors when 
  the simulations run with different cosmologies are used 
  The errors then scale with $\sqrt{n_r}$.
  The dashed horizontal line indicates the variance in the GP prior 
  imposed on each mode amplitude (where the GP prior parameters are determined by maximum-likelihood estimates given the model covariance matrices at each design point). 
}
  \label{fig:FisherErrors}
\end{figure*}

The marginal Fisher errors of the mode amplitudes for the diaginal 
and lower triagular compoments of the GCD, $\gamma$ and $\delta$ 
respectively, are
shown in Figure~\ref{fig:FisherErrors} as functions of $n_{r}$ for 
$n_d=32$. Here we compare the emulator error performance in the mode 
amplitudes to that of a brute force simulation. The errors shown are 
for a single cosmology design point, however the emulator is 
constrained over the full parameter space considered. We see that the 
emulator performs as well or better depending on the number of 
simulation run and with the added benefit of modeling the CDC over 
the full parameter space. While the fractional error at higher modes 
is constrained by the GP prior, keep in mind that the prior is 
informed from the data. The GP prior can be thought of as the maximum 
variation over all the cosmology design points.

If we compare the two lines in Fig.~\ref{fig:FisherErrors} for fixed 
fractional error (i.e. reading horizontally), we see that in many cases 
the number of simulations required for the emulator is about one 
order of magnitude smaller. 
The results for other $n_d$ values 
are similar, except that the GP prior is less significant for a fixed 
$n_r$.

We have not propagated any uncertainties from the calibration 
of the GP parameters or the truncation of the PC basis expansion.
Including these uncertainties will tend to bring the emulator forecasts 
closer to those assuming no relationship between modes at different 
design points. But, given the success of other emulator frameworks in the 
literature, we do not expect additional sources of uncertainty to 
qualitatively change the results in Fig.~\ref{fig:FisherErrors}.

\section{Conclusions}
\label{sec:conclusions}

Estimating covariance matrices of power-spectra and correlation functions
is a computationally intensive task that requires many CPU hours of simulations 
to achieve the accuracy required for future surveys
\cite{lsst09, euclid12}. In this paper we have shown that the 
precision in the estimated cosmology dependent covariance matrices 
can only improve as fast as $1/\sqrt{n_r}$, where $n_r$ is the number 
of simulation realizations, if one considers each cosmological model 
independently. 
If however, one simultaneously models the simulations at different 
cosmologies using Gaussian Processes, the number of simulations 
required to reach a given precision can be reduced while also 
modeling the CDC (which cannot otherwise be done with a sample of 
disjoint cosmological simulations). This makes the computational 
challenges of simulating the analysis much more tractable. 
However, we do not find orders of magnitude improvements in the 
number of simulations needed and other methods of estimating the 
covariances should be considered in combination with the emulator 
presented here 
(e.g. shrinkage estimators \cite{pope08}, large-scale 
mode-resampling~\cite{schneider11b}, and optimized simulation design 
spaces~\cite{schneider11a}.)

Future work will demonstrate the accuracy of the full emulator framework, with all uncertainties propagated including those from the lack of sample covariance estimators at each design point. 

\section*{Acknowledgments}
Part of this work performed under the auspices of the 
U.S. Department of Energy by Lawrence Livermore National Laboratory under Contract DE-AC52-07NA27344. Christopher Morrison acknowledges the support of NSF Grant AST-1009514.

\bibliography{covemulator}

\appendix

\section{The halo model}
\label{sec:halo_model}

For our `simulation' of covariances we employ the halo model \citep{cooraysheth02} to estimate the nonlinear matter power 
spectrum to be used in the analysis. This modeling is part of the python cosmology prediction package CHOMP\footnote{available at: http://code/google.com/p/chomp}. We use the halo model as defined in \cite{seljak00} using a \cite{st99} mass function. This model is accurate to within 20\% at 1 arcminute when compared to $N$-body simulations from \cite{sato11} and within 10\% between 20-200 arcminutes. For the covariances we follow the formalism of several papers \citep{scranton03, joachimi08, takada09}, considering two contributions to our covariance matrix, a Gaussian 
and a non-Gaussian term.

\begin{equation}
C(\theta_1, \theta_2) = C_G(\theta_1, \theta_2) + C_{NG}(\theta_i, \theta_j)
\end{equation}

For the Gaussian term we use the definitions as laid out in \cite{joachimi08} using Limber's approximation

\begin{equation}
C_G^{ijkl}(\theta_1, \theta_2) = \frac{1}{2 \pi A} \int l dl J_0(l\theta_i) J_0(l\theta_j) \{ P^{ik}(l)P^{jl}(l) + P^{il}(l)P^{jk}(l)\ + \frac{\sigma^4}{4 \bar{n}^i \bar{n}^j}(\delta_{ik}\delta_{jl} + \delta_{il}\delta_{jk})\}
\end{equation}
where $A$ is the area of the survey, $\sigma^2$ is the variance per galaxy pair, $\bar{n}^i$ is the density of galaxies for probe $i$, and $P^{ij}(l)$ is the projected power spectrum written as

\begin{equation}
P^{ij}(l) = \int_0^{\chi_H} d\chi \frac{f^i(\chi)f^j(\chi)}{\chi^{-2}}P(\frac{l}{\chi}: \chi)
\end{equation}
where $P$ is the non-linear matter power spectrum as a function of redshift, $\chi$ is the comving distance, $f^i(\chi)$ is the weighted window function (e.g. the lensing kernel), and the integration is from 0 to the horizon. 

For the non-Guassian term, we model only the one halo term of the halo-trispectrum. Following \cite{cooraywu01}, this term is

\begin{equation}
T_{1 - halo}(k_1, k_2) = \int dM \frac{dn}{dM} (\frac{M}{\bar{\rho}})^4*y(k_1, M)^2*y(k_2, M)^2
\end{equation}

Where $\frac{dn}{dM}$ iwthe number density of halos as a fucntion of mass $M$, $\bar{\rho}$ is the matter density, and $y$ is the Fourier transform of halo density profile normalized to 1 at $k=0$. For this analysis we assume that halos follow an NFW \cite{nfw97} profile. Here we only considering the tri-spectrum in a parallelogram configuration as required by the covariance estimation. Projecting this configuration with the window functions we have

\begin{equation}
\mathcal{T}^{ijkl}(l_1, l_2) = \int_0^{\chi_H} f^i(\chi)f^j(\chi)f^k(\chi)f^l(\chi)/\chi^{-6} T(\frac{l_1}{\chi}, \frac{l_2}{\chi}: \chi)
\end{equation}
The final form of the non-Gaussian covaraince is
\begin{equation}
C^{ijkl}(\theta_1, \theta_2) = \frac{1}{4 \pi^2 A} \int l_1 dl_1 \int l_2 dl_2 J_0(l_1 \theta_1) J_0(l_2 \theta_2) \mathcal{T}^{ijkl}(l_1, l_2) 
\end{equation}

It is worth noting that more terms contribute to the covariance and are highly dependent on the specified survey geometry. One parameterization as presented by \cite{th13} gives the beat coupling and halo sample variance as a single Super-Sample Covariance term. This term is as dominant as the non-Gaussian term at scales $k>1$ Mpc/h, however, we do not consider them in this analaysis currently. While this term is not sub-dominant, within the scope of this paper, we do not expect it to significantly change the covariance's dependence on cosmology or add any challenge to the marix decomposition and emulation presented in this work.

\begin{figure}
    \centerline{
        \includegraphics[width=0.5\textwidth]{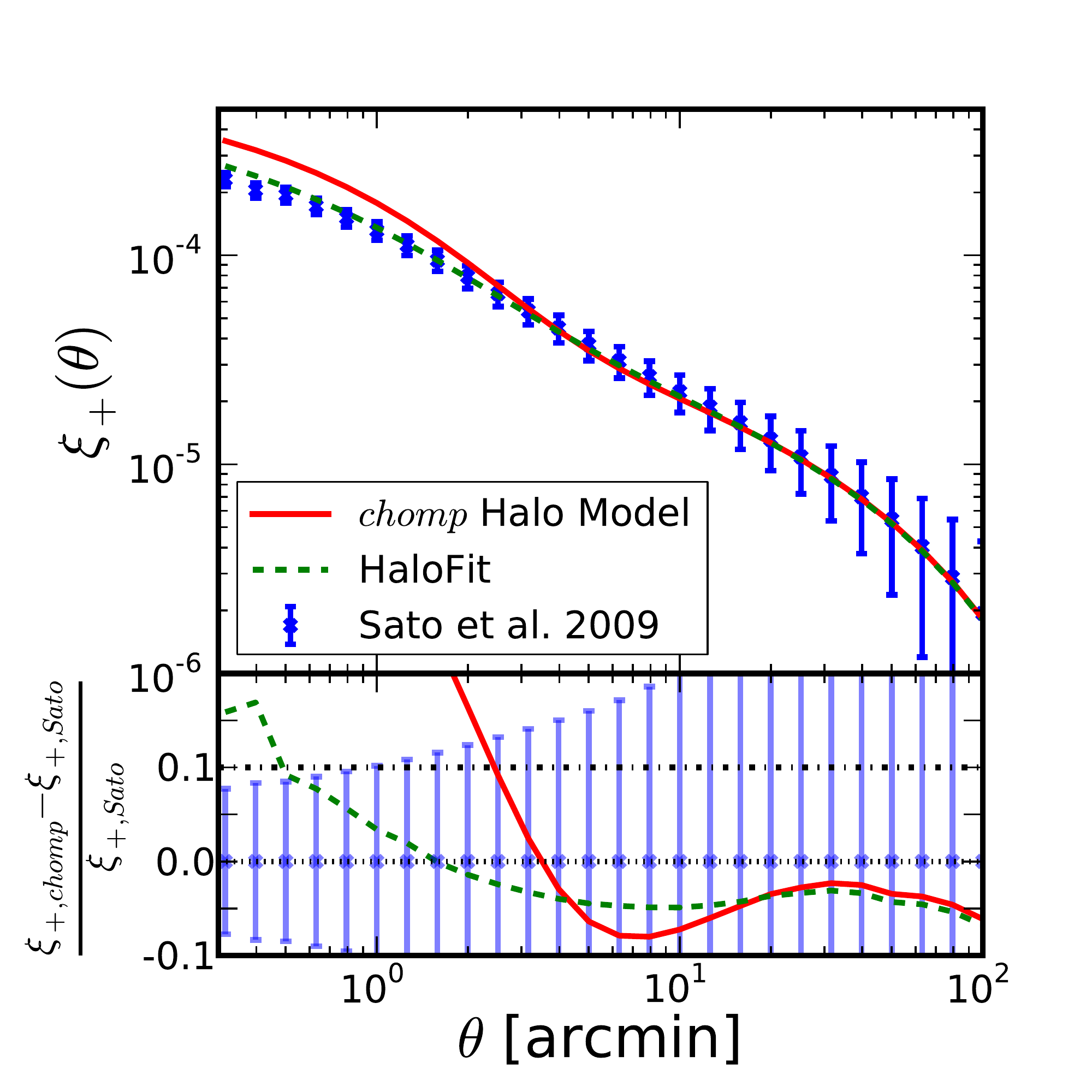}
        \includegraphics[width=0.5\textwidth]{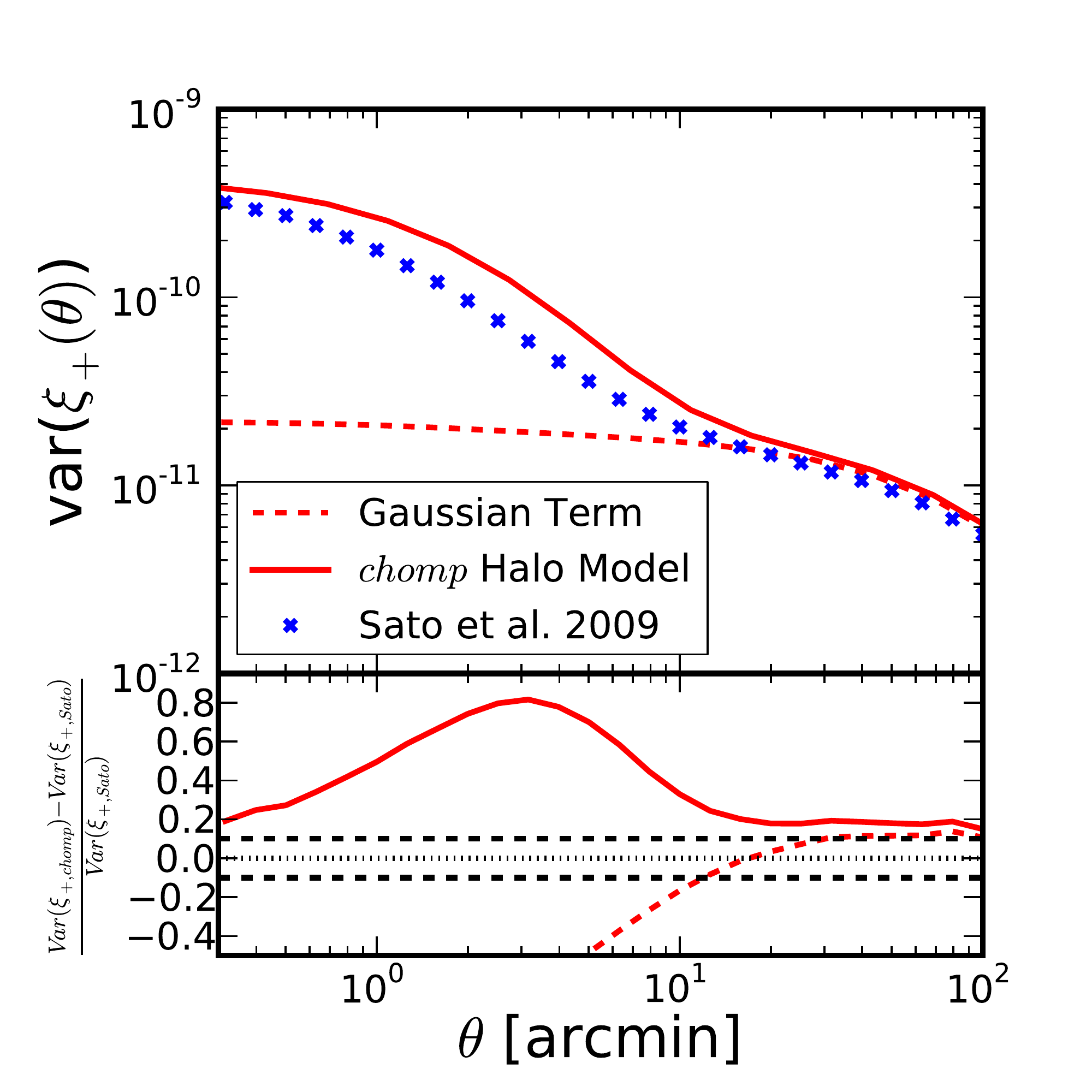}
    }
    \caption{Comparison of CHOMP lensing two-point functions with 
    those from \cite{sato09}, and \cite{sato11} using ray-tracing 
    through $N$-body simulations. Left: Comparison between the
    simulations and the $chomp$ shear correlations. Right: Comparison
    bewteen the simiulated and predicted covariance matrices. The
    correlation function is predicted to within 10\% at acminute
    scales in the halo model and to sub-arcminute scales for HaloFit.
    The covariances agree to within 20\% for scales larger than 10
    arcminutes. The results of this anlaysis hold as long as the
    variation of both the simulated and predicted covariances are
    smooth as a function of cosmology.}
    \label{fig:sato_comparison_twopoint}
\end{figure}

\section{Validation of halo model shear correlation covariance}

To apply our forecasts for planning simulation runs, 
it is important to understand how well our covariance model may
reproduce the covariance derived from simulations. 
In this section we compare our covariance model from the CHOMP code 
to the 1000 lensing simulation realizations from 
\cite{sato09, sato11}. These simulations give the power spectra and 
real-space correlations of cosmic shear for a variety of source redshifts. 
We compare our model constructed from the CHOMP cosmology package to these sims. 
The left pannel of Figure \ref{fig:sato_comparison_twopoint} shows the level of 
agreement between our models and the Sato results. We see good agreement between 
these models and the $N$-body sims. For the {\it halofit} model derived in 
\cite{takahashi12} we find agreement to within 5\% at arcminute scales for the 
real-space correlation. The Halo Model used in this analysis fare's slighly 
worse but is still within 10\% for similar scales to the HaloFit model.

The right pannel of Figure \ref{fig:sato_comparison_twopoint}
shows how well the halo model recovers the features of the Sato
covarianes. Overall, we find agreement between the simluations at various scales. The reader should keep in mind, though, that the results of this paper hold as long as the covariance matrix variation with respect to cosmology is smooth and therefore still gives an adiquate description of the covariance as a funciton of cosmology. Future work will explore the robustness and limitations of this code further.

\label{lastpage}
\end{document}